\documentclass{trbunofficial} 

\usepackage{svg}
\usepackage{newtxtext,newtxmath}
\usepackage{bm}

\usepackage{float}
\usepackage{placeins} 
\usepackage{array}  

\usepackage[hidelinks]{hyperref}

\newif\ifarxiv
\arxivtrue      

\AuthorHeaders{Ahmad, Dasgupta, Rahman, Khan, Hauque, Saba, Huynh, Zhao, and Ozguven}

\title{Lessons Learned from the Real-World Deployment of Multi-Sensor Fusion for Proactive Work Zone Safety Application}

\author{
\textbf{Minhaj Uddin Ahmad, Ph.D. Student, Corresponding Author}\\
  Department of Civil, Construction \& Environmental Engineering, The University of Alabama\\
  2009 SCIB, 28 Kirkbride Lane, Tuscaloosa, AL 35487-0288\\
  Email: mahmad12@crimson.ua.edu \\
  \hfill\break
  \textbf{Sagar Dasgupta, Ph.D.}\\
  Department of Civil, Construction \& Environmental Engineering, The University of Alabama\\
  2009 SCIB, 28 Kirkbride Lane, Tuscaloosa, AL 35487-0288\\
  Email: sdasgupta@ua.edu \\
  \hfill\break%
  \textbf{Mizanur Rahman, Ph.D.}\\
  Department of Civil, Construction \& Environmental Engineering, The University of Alabama\\
  2007 SCIB, 28 Kirkbride Lane, Tuscaloosa, AL 35487-0288\\
  Email: mizan.rahman@ua.edu \\
  \hfill\break
  \textbf{Sakib Khan, Ph.D.}\\
  Principal Intelligent Transportation Systems Engineer, MITRE Corporation\\
  7525 Colshire Dr, McLean, 22102 Virginia\\
  Email: sakibkhan@mitre.org \\
  \hfill\break
  \textbf{Md. Wasiul Haque, Ph.D. Student}\\
  Department of Civil, Construction \& Environmental Engineering, The University of Alabama\\
  2009 SCIB, 28 Kirkbride Lane, Tuscaloosa, AL 35487-0288\\
  Email: mhaque16@crimson.ua.edu\\
  \hfill\break
  \textbf{Suhala Rabab Saba, Ph.D. Student}\\
  Department of Civil, Construction \& Environmental Engineering, The University of Alabama\\
  2009 SCIB, 28 Kirkbride Lane, Tuscaloosa, AL 35487-0288\\
  Email: ssaba@crimson.ua.edu \\
  \hfill\break%
  \textbf{David Bodoh}\\
  Lead Systems Engineer, MITRE Corporation\\
  7525 Colshire Dr, McLean, 22102 Virginia\\
  Email: dbodoh@mitre.org \\
  \hfill\break
  \textbf{Nathan Huynh, Ph.D.}\\
  Department of Civil and Environmental Engineering, University of Nebraska-Lincoln\\
  262D Prem Paul Research Center at Whittier School\\
  2200 Vine Street, Lincoln 68588 Nebraska\\
  Email: nathan.huynh@unl.edu\\
  \hfill\break%
  \textbf{Li Zhao, Ph.D}\\
  Department of Civil and Environmental Engineering, University of Nebraska-Lincoln\\
  262K Prem Paul Research Center at Whittier School\\
  2200 Vine Street, Lincoln 68588 Nebraska\\
  Email: lizhao@unl.edu, (402) 472-1928\\
  \hfill\break%
  \textbf{Eren Erman Ozguven, Ph.D.}\\
  Department of Civil and Environmental Engineering, FAMU-FSU College of Engineering\\
  2525 Pottsdamer Street, Tallahassee, FL, 32311\\
  Email: eozguven@eng.famu.fsu.edu
}

\begin{document}
\maketitle
\section{Abstract}

Proactive safety systems that anticipate and mitigate traffic risks before incidents occur are increasingly recognized as essential for improving work zone safety. Unlike traditional reactive methods, these systems rely on real-time sensing, trajectory prediction, and intelligent infrastructure to detect potential hazards. Existing simulation-based studies often overlook, and real-world deployment studies rarely discuss the practical challenges associated with deploying such systems in operational settings, particularly those involving roadside infrastructure and multi-sensor integration and fusion. This study addresses that gap by presenting deployment insights and technical lessons learned from a real-world implementation of a multi-sensor safety system at an active bridge repair work zone along the N-2/US-75 corridor in Lincoln, Nebraska. The deployed system combines LiDAR, radar, and camera sensors with an edge computing platform to support multi-modal object tracking, trajectory fusion, and real-time analytics. Specifically, this study presents key lessons learned across three critical stages of deployment: (1) sensor selection and placement, (2) sensor calibration, system integration, and validation, and (3) sensor fusion. Additionally, we propose a predictive digital twin framework that leverages fused trajectory data for early conflict detection and real-time warning generation, enabling proactive safety interventions.

\hfill\break%
\noindent\textit{Keywords}: Proactive Safety, Work Zone Safety, Digital Twin, Trajectory Prediction
\newpage

\section{Background and Motivation}

Proactive safety strategies, which aim to prevent traffic incidents before they occur, are increasingly gaining traction in both research and practice. Unlike traditional reactive systems that respond after a crash or near-miss, proactive approaches rely on real-time data, predictive analytics, and intelligent infrastructure to anticipate risks and facilitate early interventions. Recent advancements in sensing, computation, and communication technologies have greatly enhanced the potential of such systems that can substantially improve traffic safety and contribute to saving lives. Roadside infrastructure equipped with heterogeneous sensor suites, such as LiDAR, radar, and camera systems, can help to develop a proactive safety system for safety-critical roadway locations, such as work zones and intersections. These sensors enable the continuous and reliable monitoring and tracking of vehicle and vulnerable road user movements across diverse lighting and weather conditions. When properly calibrated and integrated, they offer a rich and complementary data stream capable of capturing a wide range of spatial and contextual information essential for understanding, modeling, and anticipating roadway traffic behavior in real time. 

Numerous studies have explored the use of roadside infrastructure-based multi-modal sensor data to improve detection, classification, and tracking capabilities~\cite{syamal2024enhancing, wang2022object}. Different studies utilize sensor data streams at different levels of detail to support various tasks. Some fuse the information available from sensor modalities at the detection and classification level~\cite{li2024scene} while others focus on fusing vehicle trajectories generated by individual sensors~\cite{azimjonov2021real}. While simulation-based research has contributed significantly to such development, it often overlooks the complexities present in real-world deployments, such as imperfect sensor alignment, data dropout, sensor calibration, computation latency, and environmental variability~\cite{fu2020camera, drews2022deepfusion}. Real-world deployments, on the other hand, provides invaluable insights into operational constraints and challenges. These challenges include effects of terrain characteristic impacting sensor placement decisions, weather variability, stability and vibrations of the mounting pole, the height of the sensors for optimal line of sight, calibration and hardware integration issues, and system-level integration challenges that are necessary for transitioning from prototype to practice.

Studies with real-world deployments, such as \citep{li2023sensor}, \citep{zimmer2023infradet3d}, and \citep{tsaregorodtsev2024infrastructure}, have demonstrated the value of deploying and evaluating systems under real-world conditions. These deployments highlight practical limitations and design tradeoffs, such as the importance of power and network infrastructure, local storage capacity, sensor visibility constraints, and the need for accurate ground truth sources for validation. Moreover, real-world experiments often require close collaboration with public agencies and infrastructure stakeholders, introducing additional operational considerations related to safety, accessibility, and equipment protection. For instance, \citep{tsaregorodtsev2024infrastructure} deployed a real-time camera and radar-based perception system at a live urban intersection in Ulm, Germany. Their study highlighted practical challenges, such as achieving low-latency performance suitable for cooperative driving and ensuring robustness under snow and low-visibility conditions. Similarly, InfraDet3D~\cite{zimmer2023infradet3d} demonstrated a robust multi-modal 3D perception system deployed at the A9 Test Stretch in Germany using infrastructure-mounted LiDAR and camera sensors. They emphasized the benefits of fusing early and late-stage detections while introducing automatic calibration tools, HD map integration, and support for adverse lighting conditions. While these efforts underscore the feasibility of deploying infrastructure-based multi-sensor systems, most existing studies tend to focus on system performance rather than addressing the broader set of practical deployment challenges, especially those involving sensor placement, maintenance, real-time data fusion, and operational integration with existing roadside infrastructure. As a result, there remains a significant gap in the literature concerning the sustained deployment and scalability of such systems under diverse real-world conditions.


This study presents deployment insights from a real-world implementation of work zone safety applications that utilize roadside infrastructure integrated with heterogeneous sensors, including LiDAR, radar, and camera systems. Thus, specifically, this study presents key lessons learned across three critical stages of real-world deployment: (1) sensor selection and placement, (2) sensor calibration, system integration, and validation, and (3) integration of sensor-generated data into a predictive digital twin framework to support proactive safety applications. We address the challenges associated with sensor installation, calibration, and data fusion in dynamic roadside environments. Additionally, we highlight how fused vehicle trajectories can be incorporated into a predictive digital twin to enable early conflict detection and real-time warning generation.


The remainder of this paper is structured as follows. The next section describes the real-world deployment of the multi-sensor system at an active work zone, including sensor selection, placement, and site characteristics. This is followed by a discussion of key lessons learned from the deployment, with emphasis on sensor configuration, calibration challenges, and integration issues. The paper then outlines the calibration methods used for both LiDAR and radar-camera systems, including spatial alignment and coordinate transformation processes. Subsequently, we present the trajectory validation procedure using data collected from a reference vehicle equipped with a Global Navigation Satellite System (GNSS) receiver and an Inertial Measurement Unit (IMU), to evaluate the accuracy of vehicle detection, tracking, and sensor-derived trajectories. Building on these validated trajectories, the paper introduces a predictive digital twin framework that integrates fused sensor data to support proactive safety applications. Finally, we conclude with a summary of findings and a set of practical recommendations for future deployments of infrastructure-based multi-sensor systems in work zone environments.

\section{Real-world deployment of multi-sensor tracking in work zones}

\begin{figure}[!htbp]
    \centering
    \includegraphics[width=\linewidth]{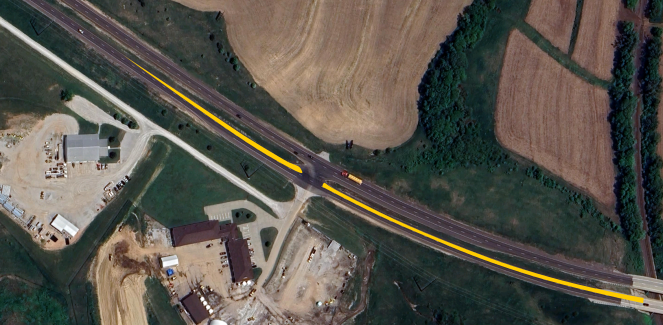}
    \caption{Satellite view of work zone along N-2/US-75 in Lincoln, NE (lane closure area highlighted in yellow).}
    \label{fig:nebraska}
\end{figure}

We established a roadside sensing deployment integrated with heterogeneous sensors, including LiDAR, radar, and camera systems, at an active bridge repair work zone site along the N-2/US-75 corridor in Lincoln, Nebraska, United States. The selected location featured dynamic traffic operations with a two-to-one lane reduction, where only a single travel lane remained open in each direction at a time. The work zone site is shown in Figure~\ref{fig:nebraska}. This configuration, combined with a temporary speed limit reduction from 70 mph to 55 mph, provided an ideal real-world environment to observe traffic flow disruptions and driver behavior under constrained conditions. The site was selected in collaboration with local transportation agencies and construction contractors following an assessment of several candidate locations along the US-75 corridor. The chosen segment exhibited operational characteristics conducive to the study of work zone mobility and safety, including frequent lane changes, vehicle merging behavior, and compliance with temporary traffic control measures. 

\begin{figure}[!htbp]
    \centering
    \includegraphics[width=\linewidth]{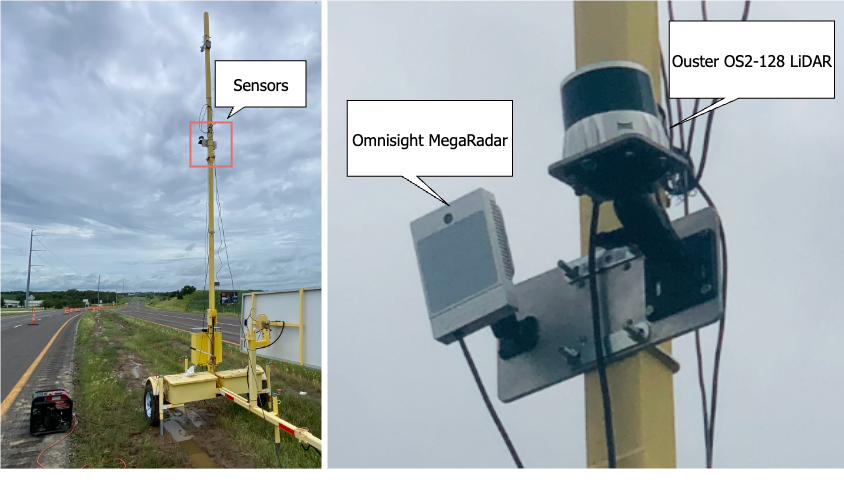}
    \caption{LiDAR and radar-camera sensor placement in the work zone.}
    \label{fig:sensors-installation}
\end{figure}

For data acquisition, two sensing platforms were deployed within the work zone: a long-range Ouster LiDAR and an integrated Omnisight radar-camera system. These systems provide multimodal tracking of vehicle movement and classification. Figure~\ref{fig:sensors-installation} illustrates the placement of the installed sensor systems within the experimental work zone. This multi-sensor configuration, comprising LiDAR, radar, and camera systems, ensures both redundancy and operational resilience. Each sensor modality has distinct strengths and limitations; for instance, cameras may struggle in low-light or foggy conditions, while LiDAR can be affected by heavy rain or snow. By fusing data from multiple heterogeneous sensors, the system can compensate for the weaknesses of any single sensor, thereby maintaining reliable performance across a wide range of environmental and operational conditions. This redundancy is particularly critical for roadside safety applications, where consistent and accurate detection of road users is essential for proactive risk mitigation. Here is a brief description of these sensor details.

LiDAR creates high-resolution 3D spatial data utilizing the reflections from its laser beams. This enables the LiDAR to construct a 3D point cloud representation with depth information. Unlike a passive sensor like a camera, LiDAR is not affected by environmental lighting, such as low light at night or lens glare. However, rain or snow may hamper the LiDAR detection as rain or snow particles disperse the laser beam. Vehicles and pedestrians can be detected, classified, and tracked by processing the LiDAR point cloud. Considerations in the choice of a particular LiDAR sensor include field of view, laser array density, detection range, and integration support by vendor-provided Software Development Kits (SDKs) to extract object-level trajectory and classification data in real-time.

Radar operates reliably under adverse weather and low-visibility conditions, such as rain, fog, or snow, utilizing reflected radio waves. It provides precise Doppler-based velocity estimations. However, the radar detections and the point cloud generated are often sparse, unlike their LiDAR counterparts. Object classification and tracking reliability may be hampered in challenging environments such as dense traffic due to sparse data points. 

Cameras can contribute to detection, classification, and tracking with rich semantic context (e.g., classification, visual cues), and are effective in daylight. However, their accuracy drops under adverse weather and low visibility conditions.

\begin{figure}[!htbp]
    \centering
    \includegraphics[width=\linewidth]{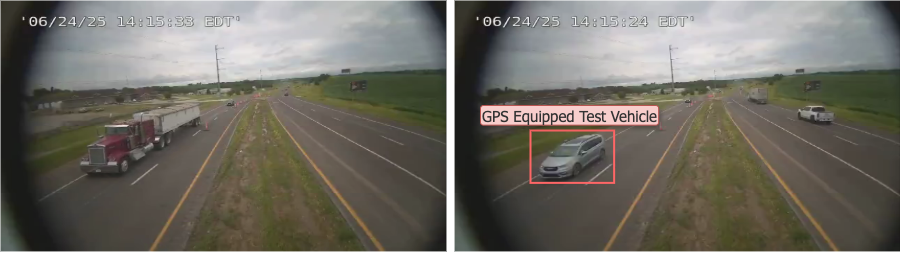}
    \caption{View of the active work zone captured by the radar-camera system.}
    \label{fig:camera}
\end{figure}

\begin{figure}[!htbp]
    \centering
    \includegraphics[width=0.5\linewidth]{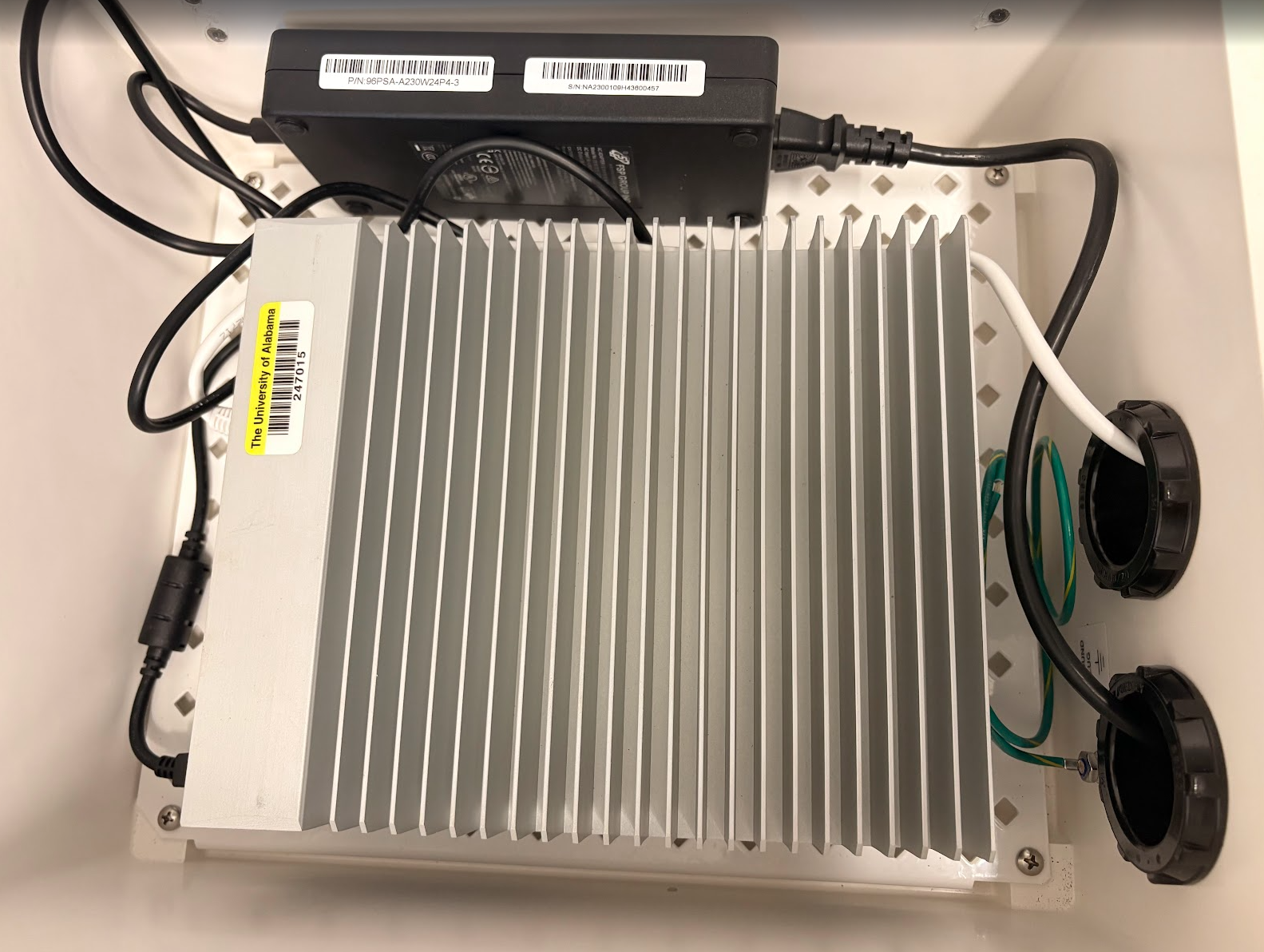}
    \caption{Edge computer for real-time processing and storing multi-sensor data stream.}
    \label{fig:edge-computer}
\end{figure}

For ground truth reference, we equipped a minivan with the NovAtel CPT7700 system, a real-time kinematic (RTK) GNSS and IMU. The Global Positioning System (GPS) is the satellite navigation system operated by the United States and represents one of several GNSS. From this point forward, we use the terms GNSS and GPS interchangeably to refer to GNSS throughout the paper. The CPT7700 integrates a single-antenna GNSS receiver with the Honeywell HG4930 MEMS IMU, offering high-accuracy positioning through combined GNSS and inertial measurements. TerraStar-L correction services were employed during the experiment, and IMU data were logged at a 10 Hz frequency to capture detailed vehicle motion, including heading and acceleration. This ground truth data served as a reference for validating sensor measurements and trajectory reconstruction algorithms. Figure~\ref{fig:camera} illustrates an image from the camera sensor, showing the work zone site from the perspective of the sensors. An edge computer, as demonstrated in Figure~\ref{fig:edge-computer}, processes and stores the incoming data from the LiDAR and radar-camera systems. The edge computer features a NVIDIA Jetson AGX Orin computing platform with 275 Terra-Flops Per Second (TOPS) processing capability and 32GB LPDDR5 RAM. It also features a 1024-core NVIDIA Ampere GPU, which enables real-time processing of dense LiDAR data.

\FloatBarrier

\section{Lessons learned}

\subsection{Selection of Sensors}

To enable real-time perception and proactive safety applications, we deploy a heterogeneous sensor suite comprising LiDAR, radar, and camera. As explained before, this multi-sensor configuration ensures redundancy and resilience, particularly in challenging conditions, such as low lighting, fog, rain, and snow, where individual sensors may underperform. To construct a digital twin of the work zone environment, the specification of data requirements is established in Table~\ref{tab:sensor_data_spec}. Different sensor modalities are often fused to complement each other to produce reliable detection, classification, and tracking. Among commercially available sensor options for roadside infrastructure, we selected sensors based on three key criteria: (1) compatibility with data specification listed in Table~\ref{tab:sensor_data_spec}, (2) long detection range suitable for highway environment, and (3) availability of vendor provided Software Development Kit (SDK) or Application Programming Interface (API) to facilitate real-time data access and integration into custom applications. In this study, the Ouster OS2-128 LiDAR and the OmniSight radar-camera fusion sensor were selected to meet these criteria. Their software support enabled data streaming to an edge computer and provided the flexibility needed to build higher-level applications such as a digital twin-based proactive safety application for work zone environments. Table~\ref{tab:sensor_specs} lists the specifications of the selected sensors. 

\begin{table}[!htbp]
\centering
\caption{Data requirement specification}
\label{tab:sensor_data_spec}
\begin{tabular}{|m{3.2cm}|m{10.8cm}|}
\hline
\textbf{Field} & \textbf{Description} \\
\hline
Timestamp & Millisecond resolution timestamps to ensure temporal alignment across modalities and consistency. \\
\hline
Track ID & Unique persistent identifier for each tracked object. \\
\hline
Classification & Object category, such as passenger vehicle, commercial motor vehicle, or pedestrian. \\
\hline
Position & Localization of each object with standard units of measurement in sensor reference frame or global reference frame. \\
\hline
Velocity & Object velocity expressed as vector components (e.g., $v_x$, $v_y$) in sensor reference frame or global reference frame. \\
\hline
\end{tabular}
\end{table}

\begin{table}[!htbp]
\centering
\caption{Technical Specifications of OmniSight FusionSensor and Ouster OS2-128 LiDAR}
\begin{tabular}{|p{4.2cm}|p{5cm}|p{5cm}|}
\hline
\textbf{Specification} & \textbf{OmniSight FusionSensor} & \textbf{Ouster OS2-128 LiDAR} \\
\hline
Sensor Type & Camera + Radar Fusion & LiDAR \\
\hline
Function & Object detection, classification, tracking & Object detection, classificatin and tracking \\
\hline
Range & 100 meter & 200 m \\
\hline
Field of View (FoV) & 120$^\circ$ &  360$^\circ$ horizontal, 22.5$^\circ$ vertical \\
\hline
Frame Rate & Up to 20 fps (fused output) & Up to 20 fps \\
\hline
Data Interface & RJ45 Ethernet & Proprietary Connector to RJ45 Ethernet \\
\hline
Points per Second & \textit{N/A} & Up to 2.6 million \\
\hline
Time Synchronization & NTP supported & PPS + NMEA / PTP supported \\
\hline
Power Consumption & 30 W PoE+ & 28 W \\
\hline
IP Rating & IP66 & IP68 / IP69K \\
\hline
Operating Temperature & –30$^\circ$C to +75$^\circ$C & –20$^\circ$C to +60$^\circ$C \\
\hline
Weight & 0.78 kg & 1.1 kg \\
\hline
Mounting & Rear-mounted bracket / plate & Bottom mount \\
\hline
\end{tabular}
\label{tab:sensor_specs}
\end{table}

\FloatBarrier

\subsection{Placement of Sensors}

In deployment preparation of the roadside sensing system, careful consideration was given to the installation height and placement of the radar-camera and LiDAR platforms to ensure optimal coverage, sensor field of view, and tracking performance. Each sensor platform comes with vendor-recommended mounting guidelines, which are informed not only by hardware constraints such as radar beam spread, camera lens properties, but also by specific sensing principles and the configuration of detection and tracking algorithms. The detection and tracking algorithms are often trained on datasets collected under specific deployment conditions, including sensor height, tilt angle, and environmental context for placement. Deviating from the vendor-recommended specifications may adversely affect detection and tracking performance by introducing discrepancies between observed data and the model's training distribution. 

The radar-camera unit (i.e., OmniSight FusionSensor), which integrates millimeter-wave radar with optical imaging, is typically recommended to be installed at an elevation of 15-20 ft from the road surface. In contrast, the Ouster LiDAR unit performs optimally at a lower mounting height of 10-15 ft, where it can generate a dense point cloud, as the laser beam tends to disperse with increasing distance from the LiDAR unit. In this study, we placed the multi-sensor system inside the work zone, in the median of the highway, ensuring the oncoming traffic is in the field of view of both the sensors. Both sensor platforms were installed at a height of 15 ft from the road surface, where both sensors could detect and track oncoming vehicles. While this configuration proved effective in our deployment, there remains potential for further improvement through additional fine-tuning of the installation height based on specific site geometry and operational objectives.

\subsection{Calibration of Sensors}

In multi-sensor deployments, each sensor type, whether it is LiDAR, radar, or camera, typically produces spatial data in its own coordinate reference system (CRS). These CRSs vary in origin, orientation, and units, depending on the sensor's design, its software and firmware, and their mounting configurations. Most LiDAR, radar, and camera systems produce spatial data of tracked objects in a local Cartesian coordinate frame centered at the sensor, while the local axes are aligned with the heading of the sensor. As such, in a multi-sensor deployment, two sensors mounted in close proximity report detected object positions in CRS frames that differ in both translation and rotation. 

To enable interpretation of the detections and tracking output produced by each of the sensors, data must be transformed into a common spatial reference. In this study, we adopt the World Geodetic System 1984 (WGS84) geodetic coordinate system as the global frame of reference. WGS84 is a widely accepted standard in navigation, mapping, and geographic information systems (GIS). GNSS receivers also produce data in the WGS84 coordinate frame. This common CRS enables compatibility to test our sensor-produced tracks with a test vehicle equipped with a GNSS receiver, as well as compatibility to plot our results on a digital map using the WGS84 coordinate system.

\begin{figure}[!htbp]
    \centering
    \includegraphics[width=0.5\linewidth]{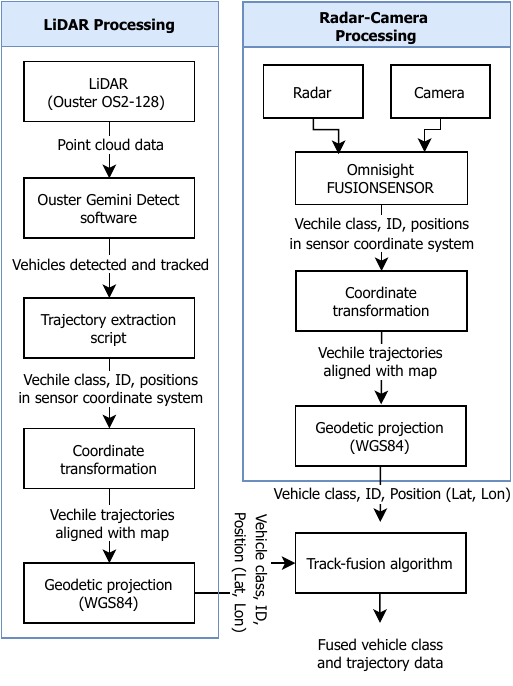}
    \caption{Real-time processing pipeline for LiDAR and radar-camera trajectory fusion.}
    \label{fig:lidar-radar-processing}
\end{figure}

We find the amount of translation and rotation that is required to transform individual sensor outputs to a common CRS through a calibration process. A sample of collected data is used for the calibration of the LiDAR and radar-camera system. Once the desired calibration values for the transformation are available, it is applied to the real-time data processing pipeline. The data processing pipeline is illustrated in Figure~\ref{fig:lidar-radar-processing}. The LiDAR and radar-camera system produces raw sensor readings that are processed by the vendor software SDK to detect and track different classes of vehicles and pedestrians. Our software client script requests and retrieves the object classification and tracking output from the vendor API (Application Programming Interface) endpoints 10 times a second (10 Hz). The corresponding coordinate transformation is applied to the spatial data produced by the sensors. Once trajectory data are in the common coordinate system, downstream track fusion algorithms can associate and fuse the individual object tracks, producing a robust vehicle trajectory. This fused trajectory can later be used by the application program. In the following subsections, we describe the calibration steps for the LiDAR and the radar-camera system. 

\FloatBarrier

\subsubsection{LiDAR Sensor Calibration}

\begin{figure}[!htbp]
    \centering
    \includegraphics[width=\linewidth]{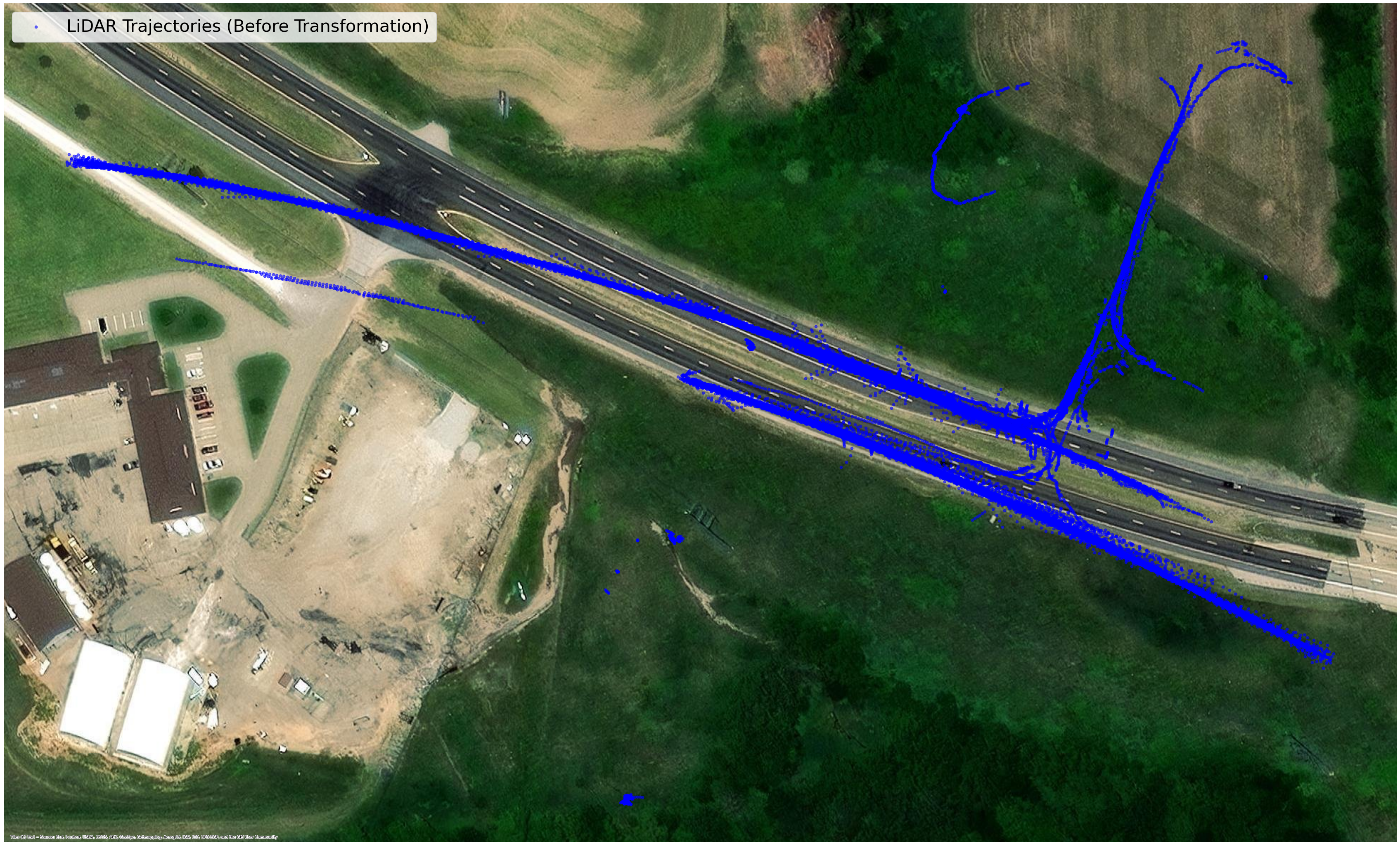}
    \caption{LiDAR trajectories overlaid on satellite map before calibration}
    \label{fig:lidar-before-correction}
\end{figure}

LiDAR sensors generate detailed spatial observation in their field of view by detecting the reflection from surrounding objects. The data is initially in point cloud format, which is processed by vendor software to produce object detections, classification, and tracking. The tracking produces a series of trajectories that are encoded in the sensor coordinate system. Using the installation site location, we create an initial visualization in the WGS84 coordinate reference frame. Figure~\ref{fig:lidar-before-correction} shows the visualization of all trajectories in the sample overlaid on ESRI satellite imagery, tracing out the shape of the road, albeit not aligned with the road. However, the geometry of the trajectories allows us to manually adjust the orientation and necessary offset such that the trajectories line up with the satellite image map. 

\begin{figure}[!htbp]
    \centering
    \includegraphics[width=\linewidth]{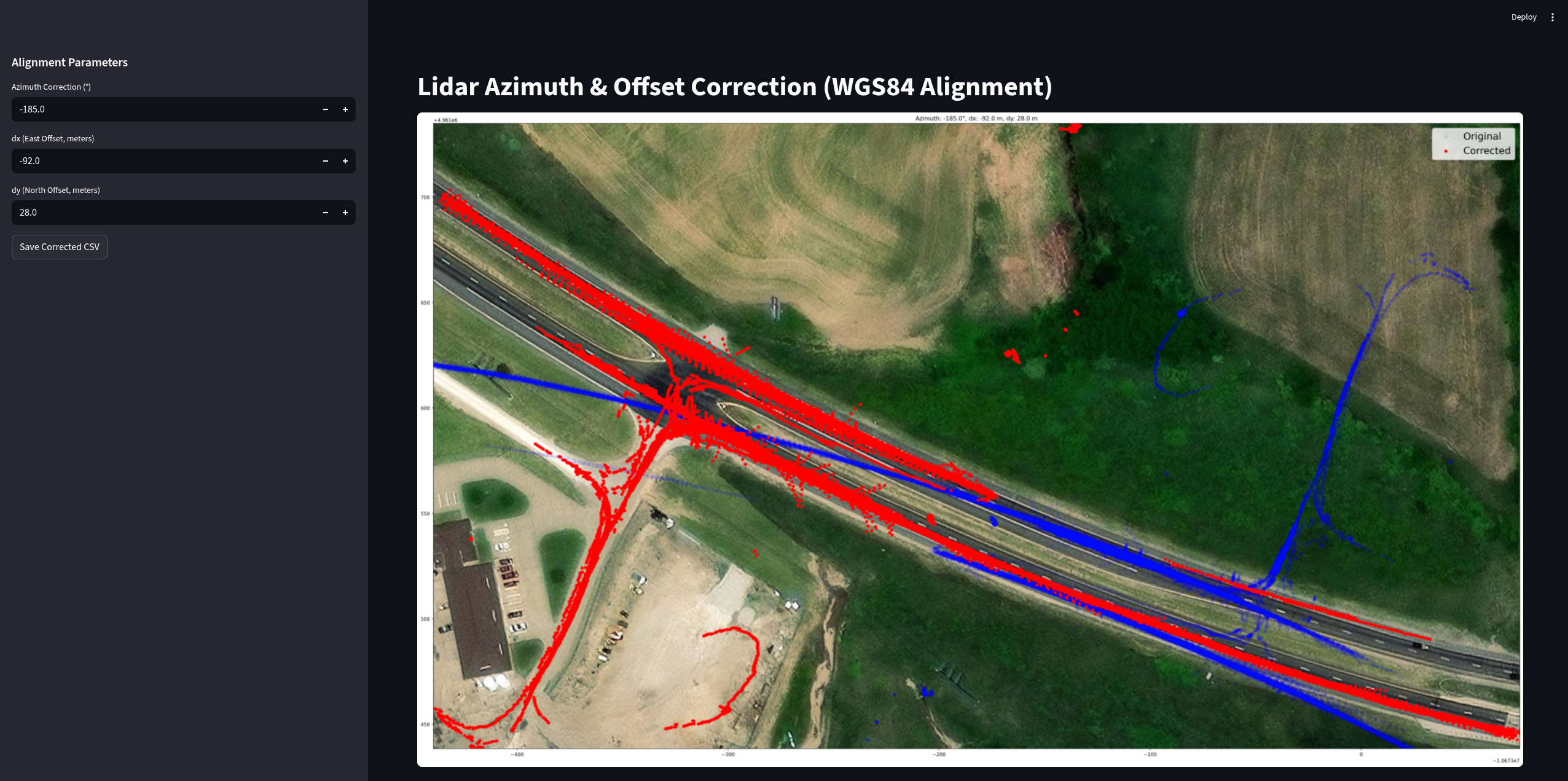}
    \caption{Interactive calibration tool for aligning LiDAR trajectories.}
    \label{fig:lidar-aligner}
\end{figure}

To support the intuitive trial-and-error alignment process, we have developed a browser-based interactive calibration tool that allows us to adjust the orientation azimuth angle and positional offsets. The interactive tool is presented in Figure~\ref{fig:lidar-aligner} and the code is provided as open-source Github Gist {\url{https://gist.github.com/minhaj6/fa3cc06bee9b6df5274133041c3cf3d4}}. The interface enables real-time visualization of the transformed trajectories over satellite imagery, facilitating manual calibration. By iteratively adjusting the transformation parameters within the tool, we manually aligned the trajectory overlay to the roadway with high visual fidelity. Once an appropriate rotation and translation were identified, we applied this transformation to the dataset to bring it into the WGS84 geodetic frame. This calibrated LiDAR data was then used in downstream fusion and analysis steps within our digital twin workflow. 

The transformation applied to each LiDAR point involved a two-dimensional similarity transformation. Let \((x, y)\) be the original coordinates in the local sensor frame, \(\theta\) the azimuth rotation angle, and \((\Delta x, \Delta y)\) the translational offsets in meters. The transformed coordinates \((x', y')\) are computed as:

\begin{linenomath}
\begin{flalign}
\begin{bmatrix} x' \\ y' \end{bmatrix}
=
\begin{bmatrix}
\cos\theta & -\sin\theta \\
\sin\theta & \cos\theta
\end{bmatrix}
\begin{bmatrix} x \\ y \end{bmatrix}
+
\begin{bmatrix} \Delta x \\ \Delta y \end{bmatrix}
\end{flalign}
\end{linenomath}\\

\begin{figure}[!htbp]
    \centering
    \includegraphics[width=\linewidth]{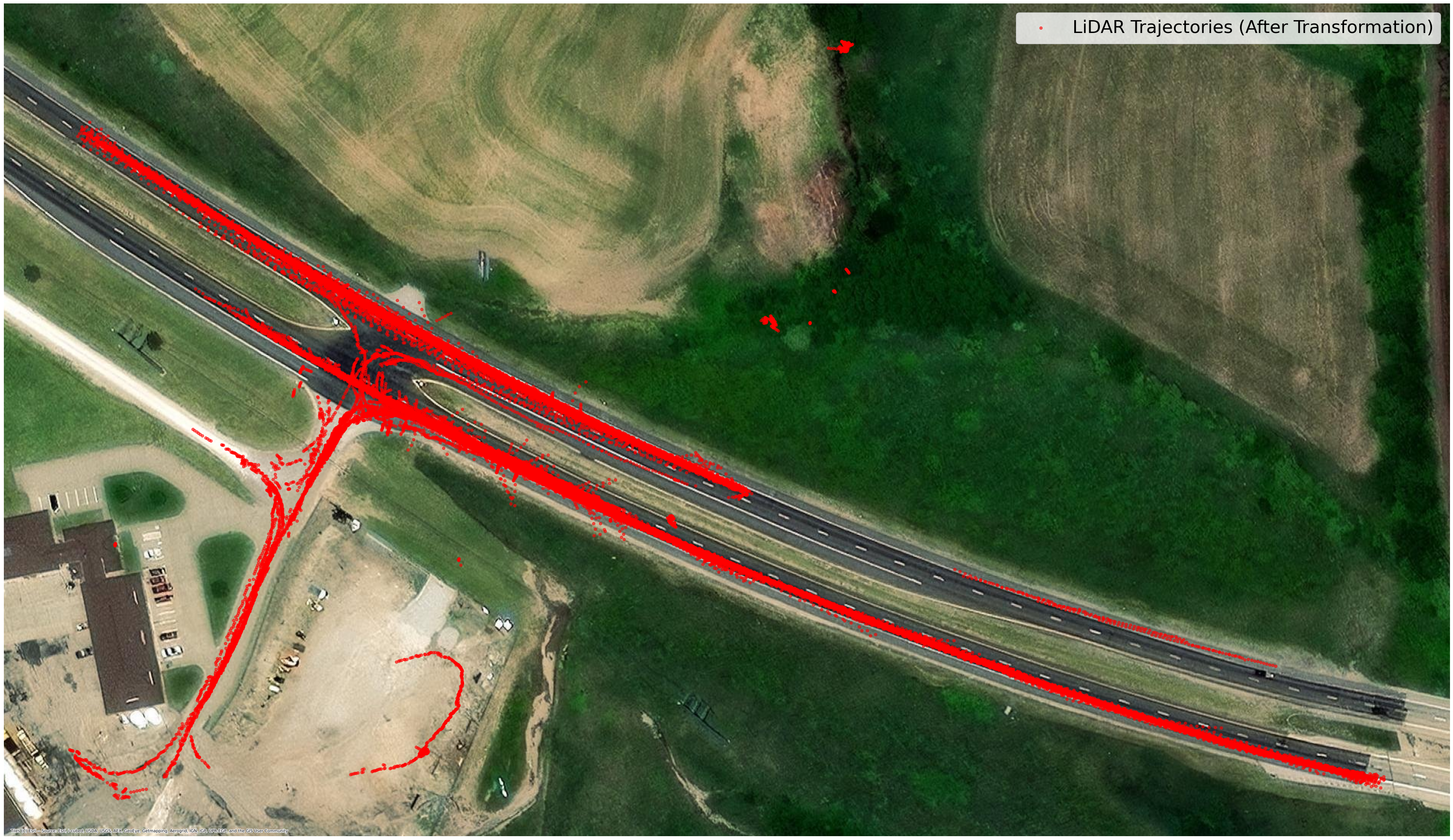}
    \caption{LiDAR trajectories overlaid on satellite map after calibration.}
    \label{fig:lidar-correction}
\end{figure}

In practice, the rotation is performed about the centroid of the trajectory cluster to preserve the overall shape. The resulting coordinates are then re-projected into the WGS84 coordinate system using an inverse map projection centered at the sensor's installation location. Figure~\ref{fig:lidar-correction} presents a collection of trajectories after the calibration process. The values found for $\theta$ are $185\circ$ from north, the north-south and east-west offsets were found to be $-92$ and $28$ meters, respectively.

\FloatBarrier


\subsubsection{Radar-Camera Sensor Calibration}

We employed a data-driven calibration approach using a rigid-body Procrustes alignment between radar-camera detected trajectories and calibrated against calibrated LiDAR detections. The calibration procedure aimed to estimate the 2D rigid transformation that is required to align the radar trajectories. The transformation comprises the necessary rotation and translation to align radar detections with the WGS84 reference frame. No scaling or shearing was assumed, as the radar detections are metric and only subject to heading and positional offsets due to sensor mounting.

The radar data used for calibration is sampled from detections and tracking produced by the OmniSight FusionSensor unit deployed at the work zone. Each detection entry in the data included a Well-Known Text (WKT)-encoded \texttt{raw\_position} field for tracked object coordinates in a local Cartesian frame. These coordinates were first extracted and converted to planar East-North coordinates using an azimuthal equidistant (AEQD) projection centered at the known sensor installation location. 

\begin{figure}[!htbp]
    \centering
    \includegraphics[width=0.85\linewidth]{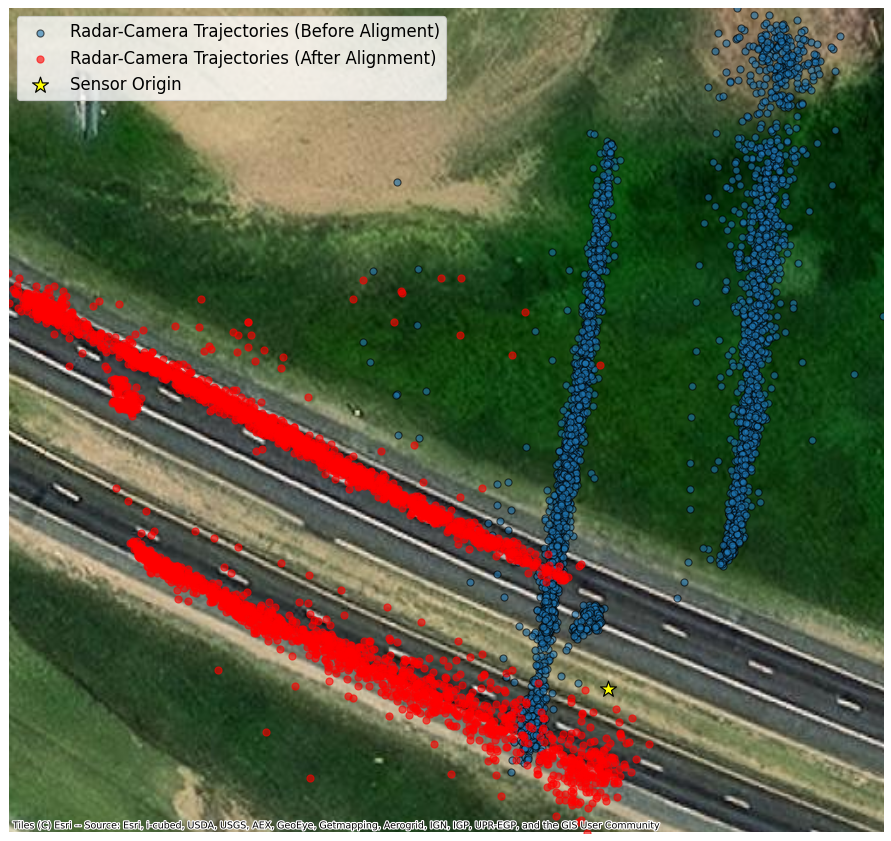}
    \caption{Radar-camera trajectories aligned using orthogonal Procrustes alignment.}
    \label{fig:radar-correction}
\end{figure}

Simultaneously, corrected LiDAR trajectories were projected into the same AEQD frame, creating two comparable point sets. To align them, we employed a Procrustes-based rigid transformation algorithm that minimizes the mean squared error between the corresponding radar and LiDAR positions. This involves computing the optimal rotation matrix $R$ and translation vector $t$ minimizing the Frobenius norm $|RA + t - B|_F$, subject to the constraint $R^\top R = I$ and $\det(R) = 1$. The solution was obtained by first mean-centering both radar-camera trajectory $A \in \mathbb{R}^{n \times 2}$ and LiDAR trajectory $B \in \mathbb{R}^{n \times 2}$, computing the cross-covariance matrix $H = (A - \bar{A})^\top (B - \bar{B})$, and applying singular value decomposition $H = U \Sigma V^\top$. The optimal rotation was $R = VU^\top$, with a reflection correction applied if $\det(R) < 0$. The translation was then computed as $t = \bar{B} - R\bar{A}$.

After solving for $R$ and $t$, we applied the resulting transform to each radar trajectory candidate and evaluated the alignment quality. The radar ID with the strongest spatial correspondence to the LiDAR trajectory—based on time-synchronized matching and shape similarity—was selected. In our deployment, the radar ID \texttt{5fd4e80b-f1b1-4418-b0c5-e3f0d9dba021} was identified automatically based on this criterion. The derived sensor yaw angle was $-109.74\circ$, representing the rotation needed to align the radar’s internal coordinate frame with the north–aligned AEQD reference frame. This yaw captures the heading offset of the radar’s axes from geographic north and is independent of any particular vehicle’s trajectory. The translation vector accounts for the sensor’s spatial offset from the origin of the AEQD projection. The final transformation was applied to all radar detections, enabling consistent fusion with LiDAR and GPS data for downstream processing in the digital twin framework. The aligned radar trajectory was then projected back into WGS84 coordinates and visualized over a satellite basemap, illustrated in Figure~\ref{fig:radar-correction}. The spatial alignment showed strong agreement with LiDAR observations. Using this value of azimuth angle and the sensor installation location as origin, the calibration can be integrated into the real-time processing pipeline.

\FloatBarrier

\subsection{System validation}

\begin{figure}[!htbp]
    \centering
    \includegraphics[width=\linewidth]{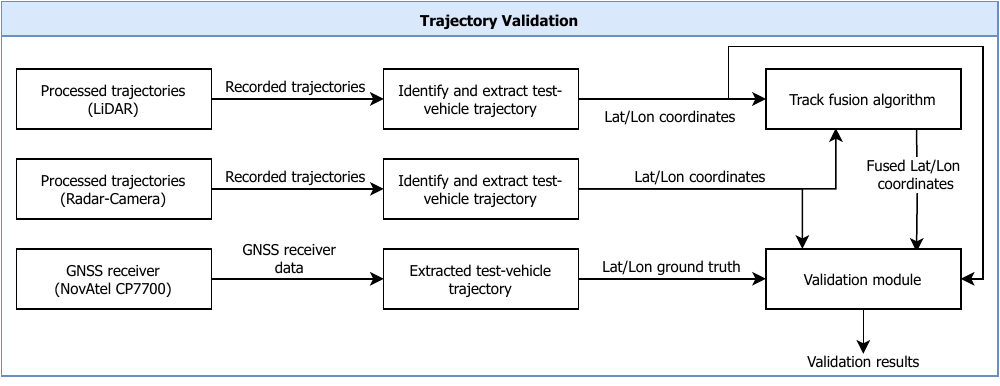}
    \caption{Trajectory validation workflow comparing sensor tracks with GNSS.}
    \label{fig:trajectory-validation}
\end{figure}

Figure~\ref{fig:trajectory-validation} illustrates the workflow designed for validating the trajectories produced by the LiDAR and radar-camera sensors. From the processed and correctly aligned sensor trajectories, we identify and extract the test vehicle trajectory by filtering the time window when GNSS-equipped vehicles are in the sensor's field of view, as well as vehicle classification and the geometric shape of the trajectory from the data. The test vehicle trajectory is further validated visually using the recorded video from the camera.  

\begin{figure}[!htbp]
    \centering
    \includegraphics[width=1\linewidth]{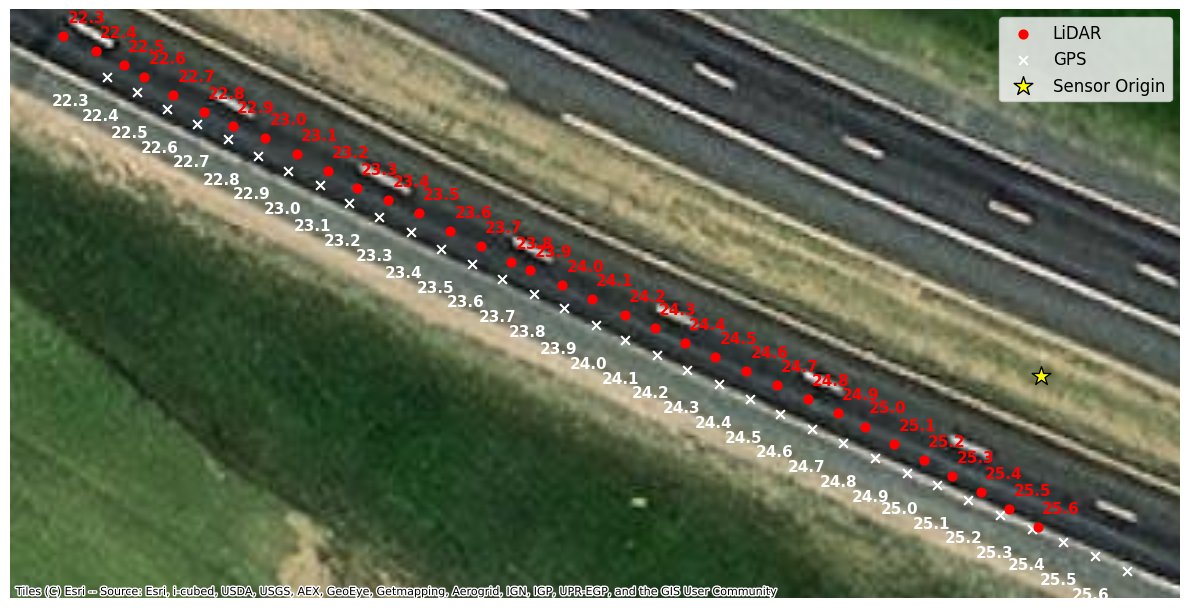}
    \caption{LiDAR trajectory of test vehicle versus GPS trajectory showing LiDAR tracking capability and alignment.}
    \label{fig:lidar-gps}
\end{figure}

In the LiDAR data sample, vehicle ID 5769 was identified as the test vehicle. Figure~\ref{fig:lidar-gps} presents the LiDAR trajectory visualization of the test vehicle along with the RTK GNSS-derived trajectory, along with their respective timestamps. The timestamps annotated in the figure represent only the seconds and milliseconds component, as this level of temporal resolution is most relevant for observing fine-scale alignment and delays. The visualization demonstrates lane-level accurate tracking of the test vehicle by the LiDAR. A constant time-lag in the LiDAR trajectory track is observed, which can be explained by the clock source of the different logs. The GNSS trajectory log uses time from GNSS receivers, which is precise time information derived from GNSS signals. However, the LiDAR logging utilizes the Network Time Protocol (NTP) server, which can add errors. We observe approximately 200 milliseconds delay, which will be accounted for in our future testing by equipping the edge computer with a GNSS clock source.

\begin{figure}[!htbp]
    \centering
    \includegraphics[width=1\linewidth]{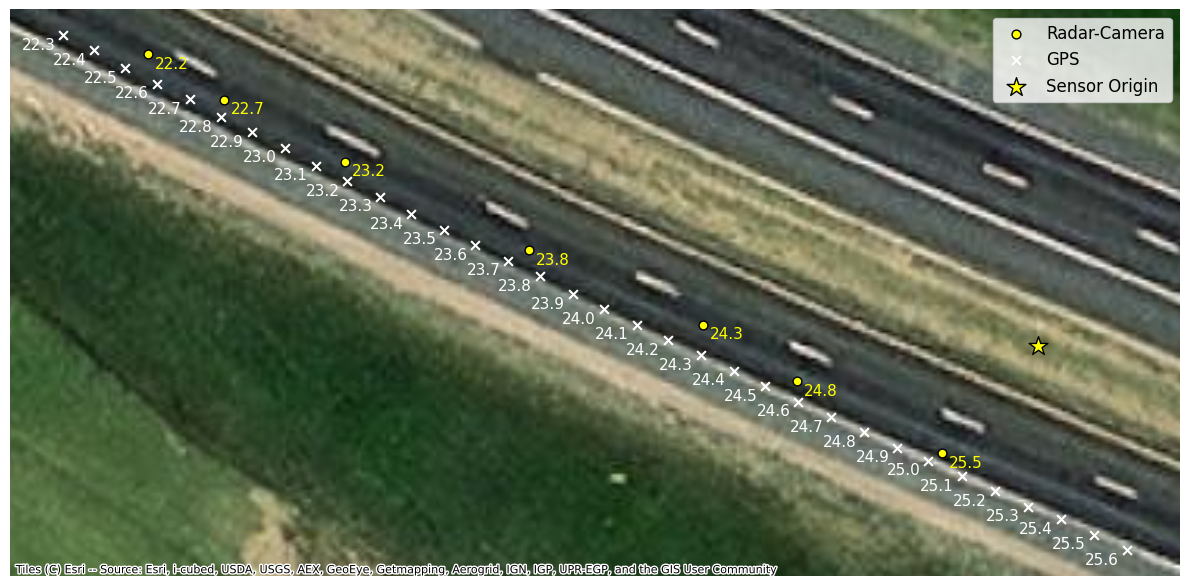}
    \caption{Radar-camera trajectory of test vehicle versus GPS trajectory showing radar-camera tracking capability and alignment.}
    \label{fig:radar-gps}
\end{figure}

Similarly, Figure~\ref{fig:radar-gps} presents the trajectory of the identified test vehicle ID \texttt{5fd4e80b\allowbreak -f1b1\allowbreak -4418\allowbreak -b0c5\allowbreak -e3f0d9dba021} along with RTK GNSS trajectory, demonstrating a similar outcome. We observe sparse data points in the radar-camera generated trajectory; however, the trajectory shows lane-level accuracy. 

\begin{figure}[!htbp]
    \centering
    \includegraphics[width=1\linewidth]{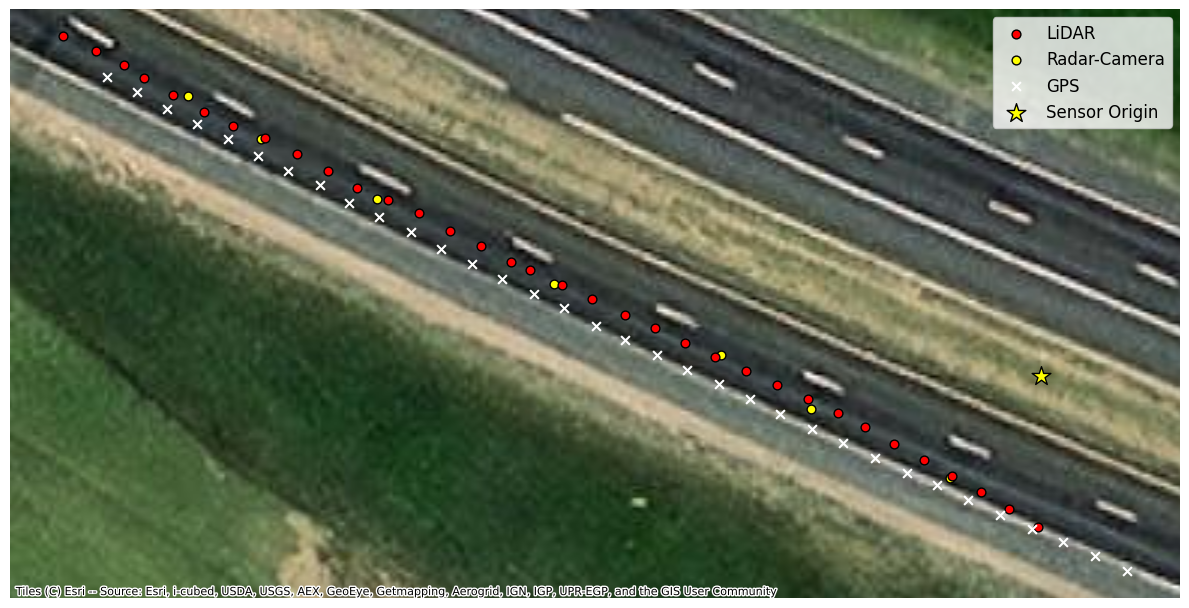}
    \caption{Overlay of LiDAR, radar-camera, and GPS trajectories for the test vehicle, demonstrating consistency across sensor modalities.}
    \label{fig:lidar-radar-gps}
\end{figure}

Figure~\ref{fig:lidar-radar-gps} presents trajectories from both sensor systems along with the GNSS trajectory for a comparative visualization.

\begin{figure}[!htbp]
    \centering
    \includegraphics[width=1\linewidth]{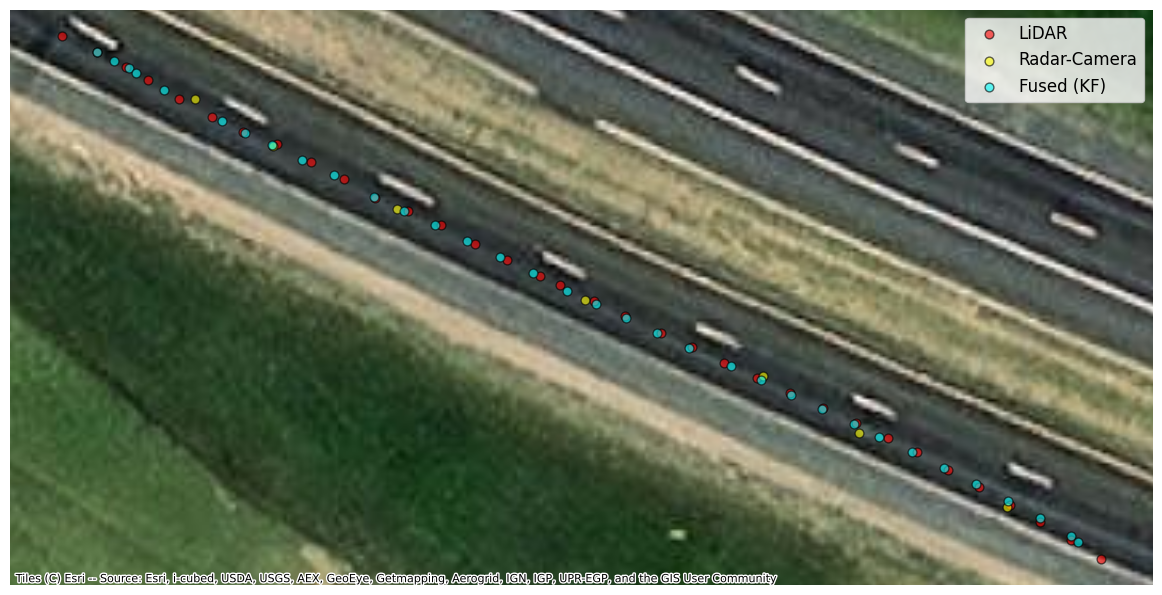}
    \caption{Fused trajectory demonstrating continuity and robustness against missing sensor data.}
    \label{fig:kalman-fused}
\end{figure}

For real-world implementation, since detections and tracking from all sensors may not be available for all timestamps, due to the inherent limitations of the technology, it is necessary to fuse the sensor data with a robust state estimation algorithm that produces a reliable trajectory by fusing the sensor data. Figure~\ref{fig:kalman-fused} demonstrates a fused trajectory utilizing a Kalman Filter (KF) that reliably combines the sensor-generated tracks using measurement updates when data is available from each sensor. This approach ensures a robust trajectory generation even when one of the sensors may fail to detect, and ensures a more reliable trajectory for downstream applications to utilize.  

\FloatBarrier

\subsection{Application of multi-sensor tracking and fusion for digital twin-based safety application}

Digital Twin (DT) technology offers a compelling framework for integrating roadside infrastructure-based detection, tracking, and predictive analytics for proactive safety applications in transportation environments. A DT is a dynamic, virtual representation of the physical world that is continuously synchronized with real-time sensor data. This bidirectional data flow between the physical and virtual layers allows the system to accurately reflect the current state of roadway assets, vehicles, workers, and environmental conditions. In the context of intelligent transportation systems, Transportation Digital Twins (TDTs) extend this concept by focusing specifically on modeling and monitoring traffic flow, infrastructure conditions, and multi-agent interactions on roadways, thereby enabling advanced safety-critical applications such as real-time incident detection, risk evaluation, and trajectory forecasting~\cite{irfan2024towardstdt,10616167}.

A particularly promising extension of the TDT paradigm is the Predictive Digital Twin (PDT). PDT systems integrate real-time sensor feeds, historical trajectory data, and machine learning models with physical simulation engines to forecast future states of a transportation environment. These forecasts simulate multiple plausible future scenarios involving connected vehicles, roadside workers, and heavy machinery. By continuously evaluating the safety implications of predicted behaviors, PDT systems can identify hazards before they materialize—such as near-collisions, dynamic geofence violations, or unsafe worker movement. This anticipatory capability enables a shift from reactive response to proactive safety enforcement, improving both operational efficiency and protection for all road users.

\begin{figure}[!htbp]
\centering
\includegraphics[width=0.85\linewidth]{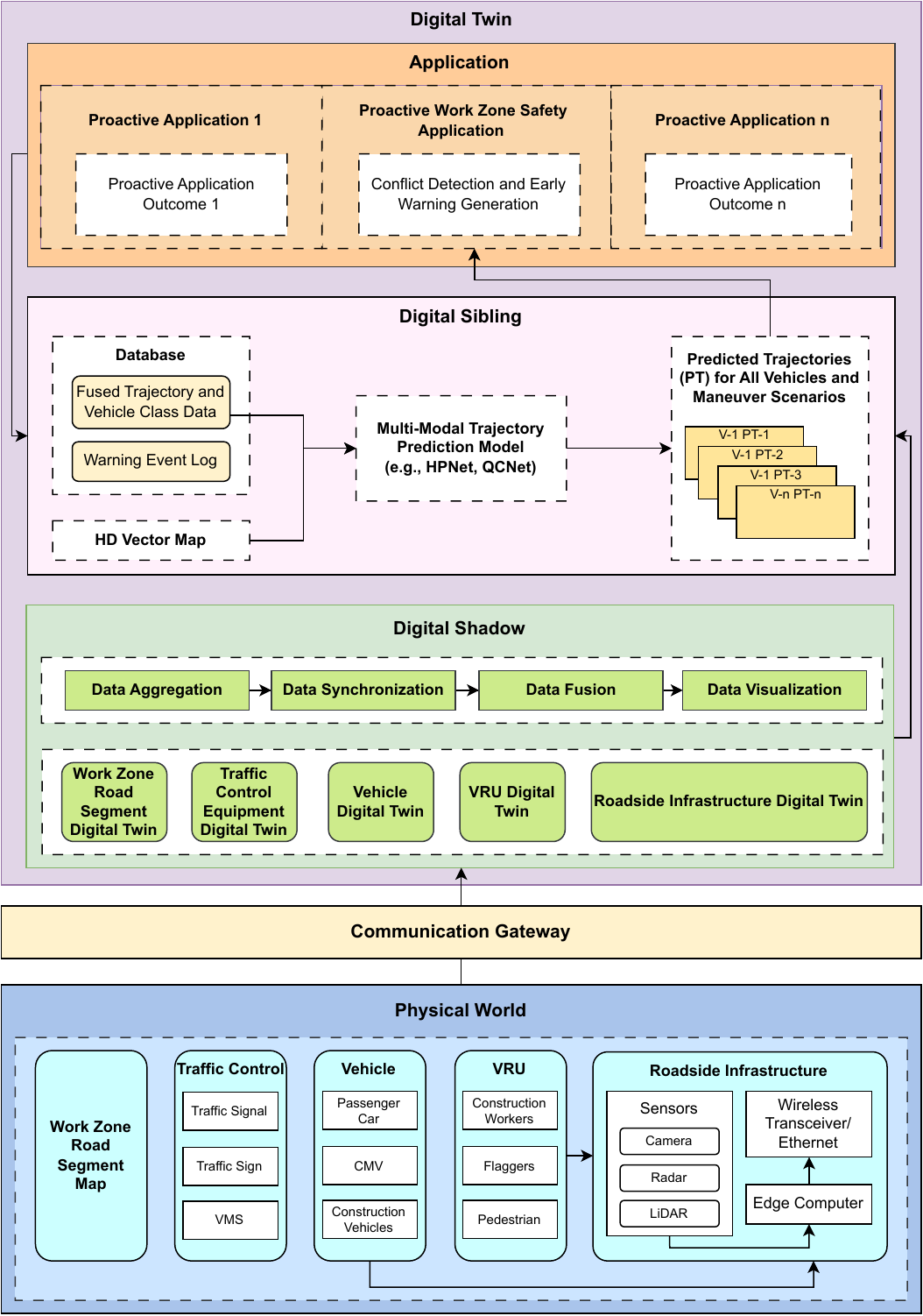}
\caption{Predictive Digital Twin framework for proactive work zone safety applications.}
\label{fig:digital-twin}
\end{figure}

\FloatBarrier

Figure~\ref{fig:digital-twin} illustrates the architecture of a real-world DT system tailored for work zone safety. The system is organized into two tightly integrated domains: the physical world and its digital twin counterpart. In the physical domain, the work zone road segment is equipped with roadside sensors, including LiDAR, radar, and cameras, which monitor the movement and behavior of construction workers, flaggers, pedestrians, passenger cars, construction vehicles, and commercial motor vehicles (CMVs). These sensors are connected to an edge computing unit via wireless or wired networks, enabling low-latency, high-frequency data transmission. Additional components such as traffic signals, variable message signs (VMS), and static signage are also monitored and mirrored in the digital twin.

The incoming multi-modal sensor data undergoes aggregation, synchronization, and fusion to form the real-time "Digital Shadow." This layer maintains an accurate, high-resolution spatial and semantic representation of all agents and infrastructure. It includes fused trajectories, object classifications, and alignment with high-definition (HD) vector maps. On top of this, the “Digital Sibling” layer enables predictive simulation. Within this layer, multi-modal trajectory prediction models forecast the future motion of each detected agent under various plausible maneuvering scenarios. These predictions, along with metadata such as object class, predicted intent, and confidence scores, are stored and visualized in the system’s central database.

The computational engine within the Digital Sibling can simulate each entity’s motion using a combination of physical models and data-driven approaches. This allows the system to generate diverse maneuver possibilities for each actor and evaluate them for spatial-temporal conflicts. Key safety metrics—such as Time-to-Collision (TTC), Post-Encroachment Time (PET), and minimum predicted distance—are computed for each trajectory pair. These risk indicators can be feed into a risk evaluation framework that ranks scenarios by severity and urgency.

When high-risk interactions can be detected—such as a pedestrian entering a vehicle’s projected path or two vehicles approaching a collision course—the system activates safety interventions. These include real-time alerts to drivers via Basic Safety Messages (BSMs) or Personal Safety Messages (PSMs) transmitted over C-V2X, and audio or haptic alerts to workers through wearable devices. In cases of repeated or extreme violations, the system can also initiate virtual enforcement mechanisms. Dynamic safety geofences can be continuously updated based on the latest predictions, and the system can also trigger rerouting logic or command enforcement alerts depending on context. All alerts and events are recorded in a structured log for post-analysis and system refinement.

The framework supports the simultaneous execution of multiple proactive safety applications. These include dynamic worker protection zones, predictive driver warning systems, real-time conflict detection, and adaptive signal control in high-risk areas. A visualization dashboard enables traffic management personnel to monitor live and predicted system states, including risk levels and system responses. Additionally, the recorded event logs can inform long-term enforcement strategies, safety audits, and system performance evaluation.

This PDT architecture enables the immediate deployment of a broad range of proactive safety capabilities. Through predictive modeling and real-time fusion of sensor data, the system can anticipate and mitigate risks such as vehicle-to-worker conflicts or dangerous vehicle interactions. It supports V2X-based driver alerts using BSMs or SPaT/MAP messages over PC5 or Uu interfaces, providing advance notice of stopped vehicles, active workers, or speed reductions. Real-time Vulnerable Road User (VRU) trajectories enable dynamic geofencing that alerts oncoming traffic when safety thresholds are violated. Meanwhile, operators gain high-resolution visibility into work zone dynamics via interactive dashboards. The architecture also facilitates proactive enforcement by tracking repeated violations in the warning event log. Finally, by modeling the real-time status of traffic control infrastructure, the DT system can support adaptive signal control algorithms that prioritize worker and VRU safety during critical phases. Together, these components create a comprehensive, intelligent, and scalable safety system that shifts the paradigm from passive monitoring to proactive protection.

\section{Conclusion and Recommendations}
This study presented practical insights from the real-world deployment of a multi-sensor vehicle tracking system designed for proactive safety applications in a work zone setting. Drawing from hands-on implementation experience, we addressed the technical and operational challenges of deploying LiDAR, radar, and camera sensors integrated with edge computing infrastructure. Specific issues related to sensor placement, calibration, time synchronization, and data fusion were discussed in detail. Through extensive calibration procedures and ground truth validation using RTK-GNSS and an IMU-equipped reference vehicle, we demonstrated the system’s ability to generate lane-level accurate, fused trajectories and robust tracking for downstream applications. We also explored how these fused trajectories can feed into predictive digital twin frameworks, enabling proactive conflict detection and early warning generation. Future deployments can benefit significantly from careful sensor placement, synchronization, and integration strategies, thereby further advancing proactive safety in dynamic transportation environments that extend beyond work zones.

Based on our observations from the real-world multi-sensor deployment at the Nebraska work zone site, several practical recommendations emerged to enhance detection and tracking performance, improve spatial coverage, and increase system reliability. These insights can inform future deployments of infrastructure-based proactive safety systems:

\begin{itemize}
    \item \textbf{Sensor Placement Considerations}: It is beneficial to install sensors on stable, elevated platforms with unobstructed lines of sight to reduce occlusions. Considering roadway geometry, grade, and curvature during placement could help improve overall sensor coverage and tracking performance. The range of the sensor system may be increased to satisfy application requirements by placing sensors in more than one location. This could be particularly beneficial where an unobstructed line of sight is short due to sharp curves and hills. 
    \item \textbf{Time Synchronization}: To support accurate data fusion across sensor types, equipping edge computing units with GNSS-synchronized clocks may help address latency and synchronization challenges.
    \item \textbf{Standardized and Modular Application Development}: Adopting open and standardized data protocols can facilitate interoperability across diverse sensor systems, such as replacing the LiDAR, radar, and camera sensors from different vendors with little friction. Leveraging a flexible architecture, such as a DT framework, can streamline integration, support future scalability, and simplify system maintenance. This approach enables modular application development, where new safety applications can be added, updated, or maintained independently without disrupting the overall system. It also facilitates long-term insight generation by organizing and preserving historical data for analysis.
    \item \textbf{Sensor Calibration}: Multi-sensor calibration process can potentially be streamlined using an automated optimization technique if a known GNSS-equipped test vehicle can be uniquely identified automatically in individual sensor data. Establishing periodic recalibration intervals can help maintain sensor accuracy over extended period of deployments, accounting for environmental variations, and physical drift of sensor hardware due to high wind or vibrations. Clear documentation of calibration procedures might also ensure consistent system performance across deployment sites. 
    \item \textbf{Sensor Fusion}: Fusion of trajectories across sensor modalities using state estimation algorithms, including Extended Kalman Filter (EKF) and Unscented Kalman Filters (UKF), can effectively handle intermittent or noisy sensor data, thereby improving trajectory reliability. Well-calibrated state estimation algorithms with sensor noise and process noise modeling can further enhance the quality of the fused trajectory. 
    \item \textbf{Stakeholder Involvement and Coordination}: Active communication with stakeholders such as transportation agencies, construction contractors, technology vendors, and local authorities from an early stage of the project can help identify site-specific constraints and requirements while minimizing operational frictions of deployment.
    
\end{itemize}

\section{Acknowledgments}
This study was financially supported by the Federal Motor Carrier Safety Administration (FMCSA) of the United States Department of Transportation (US DOT). The views expressed in this paper are solely those of the authors, who are responsible for the accuracy and factual content presented. The contents do not necessarily reflect the official views or policies of the FMCSA and US DOT.

We have used `ChatGPT4o' to rephrase some of our own writing for editorial purposes.

\section{Authors Contribution}
\textbf{Minhaj Ahmad, Sagar Dasgupta, Mizanur Rahman:} conceptualization, methodology, data collection, data analysis, and writing – original draft; \textbf{Sakib Khan:} conceptualization, data collection, writing – review and editing, and funding acquisition; \textbf{Md. Wasiul Haque, Suhala Rabab Saba, and David Bodoh:} data collection and writing – review and editing; \textbf{Mizanur Rahman, Sakib Khan, Li Zhao, Nathan Huynh, and Eren Erman Ozguven:} conceptualization, writing – review and editing, and funding acquisition.

\newpage

\bibliographystyle{trb}
\bibliography{main}
\end{document}